\documentclass[RNAAS]{aastex62}

\graphicspath{{./}{figures/}}

\begin{document}

\title{The 2018 Census of Luminous Blue Variables in the Local Group}


\correspondingauthor{Noel Richardson}
\email{noel.richardson@utoledo.edu}

\author[0000-0002-2806-9339]{Noel D. Richardson}
\affiliation{Ritter Observatory, Department of Physics and Astronomy, The University of Toledo, Toledo, OH 43606-3390, USA}

\author[0000-0002-9564-3302]{Andrea Mehner}
\affiliation{ESO - European Organisation for Astronomical Research in the Southern Hemisphere, Alonso de Cordova 3107, Vitacura, Santiago de Chile, Chile}


\section{} 

Luminous Blue Variables (LBVs, S Doradus variables, Hubble-Sandage variables) are evolved massive stars undergoing dramatic photometric and spectroscopic variations \citep{HD94}. Their general properties include high luminosities, significant mass-loss rates, variability on multiple time-scales, eruptive events, and often nebulous ejecta. Bona fide LBVs exhibit variability patterns that may take years to decades to observe. Therefore, many objects are only considered candidate LBVs (cLBVs). It is still unclear how LBVs fit into the evolutionary sequence for massive stars. The underlying physical trigger(s) of the instability is still debated. Recent research shows that LBVs could be the product of binary evolution \citep{2015MNRAS.447..598S,2016MNRAS.461.3353S}, although this has been controversial \citep{2016ApJ...825...64H}. This research note compiles confirmed and candidate LBVs in the Milky Way and Local Group galaxies.


Table \ref{table} presents the 2018 census of (c)LBVs in the Local Group. When available, popular naming designations are given in the first two columns. Confirmed LBVs are highlighted in boldface. The third column lists the 2MASS designations, and hence coordinates, labeled with a `J'. ICRS coordinates are provided in cases where no 2MASS designation exist. We present optical $V$-band and near-infrared $JHK$-band magnitudes in columns 4--7, with values taken from SIMBAD \citep{2000AaAS..143....9W}, 2MASS \citep{2006AJ....131.1163S}, \citet{2007ApJ...659.1241R}, \citet{2014ApJ...790...48H}, or \citet{2015MNRAS.447.2459S}. The final column lists relevant citations. In many cases, the objects are studied extensively and we made a reasonable choice for the reference. One major source of discovery for (c)LBVs beyond the Galaxy and Magellanic Clouds comes from the Local Group Galaxies Survey \citep{2007AJ....134.2474M,2016AJ....152...62M}. A criterion on the cLBV membership in this census is spectroscopic confirmation for the stars to be hot or P Cygni-like. With the continuing work of the {\it Gaia} mission, we expect that the region of the upper Hertzsprung-Russell diagram where LBVs reside will be better constrained in the coming years  \citep[][]{2018arXiv180503298S}. 

This census includes a few notable stars that are either LBVs or related objects. For example, HD 5980 underwent an LBV-like eruption and has long-term trends in its light curve that make it similar to LBVs \citep{2010AJ....139.2600K}. Romano's star shows long-term variability with a spectrum changing from a B star to an WN8 spectrum \citep{2016AJ....151..149P}. $\eta$ Carinae is unusual compared to other LBVs \citep[][]{2012ASSL..384.....D}.


\startlongtable
\begin{longrotatetable}
\begin{deluxetable}{lccccccc}
\tablecaption{Luminous Blue Variables in the Local Group. \label{table}}
\tablehead{
\colhead{} 		& \colhead{Alternative} & \colhead{2MASS ID} 	& \colhead{$V$} 	& \colhead{$J$} 	& \colhead{$H$} 	& \colhead{$K$} 	& \colhead{}   \\
\colhead{Name} 	& \colhead{Designation} & \colhead{or coords.} 	& \colhead{(mag)} 	& \colhead{(mag)} 	& \colhead{(mag)} 	& \colhead{(mag)} 	& \colhead{Reference} \\
}
\startdata
 \multicolumn{8}{c}{\bf Galactic LBVs}\\ \hline
  HD 80077 			&	Pismis 11 1		&	J09155478$-$4958243		&	9.00		&	4.44		&	3.93		&	3.63		&	vG01	\\
{\bf HD 90177}		&	HR Carinae		&	J10225385$-$5937283		&	9.07		&	4.56		&	3.89		&	3.64		&	HD94	\\
{\bf $\eta$ Carinae} &	HD 93308		&	J10450360$-$5941040		&	 4.5		&	2.1			&	1.3			&	0.4			&	HD94	\\
Hen 3-519			&	WR 31a			&	J10535958$-$6026444		&	10.85		&	7.32		&	6.72		&	6.10		&	vG01	\\
{\bf AG Carinae}	&	HD 94910		&	J10561158$-$6027128		&	6.96		&	5.42		&	5.08		&	4.53		&	HD94	\\
{\bf WRAY 15-751}	&	V432 Car		&	J11084006$-$6042516		&	11.79		&	7.96		&	7.24		&	6.75		&	vG01	\\
Sher 25				&	\ldots			&	1115078$-$611517		&	12.27		&	8.61		&	8.03		&	7.72		&	C05		\\
{\bf {[}GKM2012{[} WS1}	&	\ldots		& 	J13362862$-$6345387		&	15.31		&	10.33		&	9.48		&	8.90		&	\citet{2015MNRAS.449L..60K}	\\
{\bf WRAY 16-137}	&	IRAS13467-6134	&	J13501536$-$6148552		&	15.5		&	6.55		&	5.34		&	4.41		&	\citet{2014MNRAS.445L..84G}		\\
{[}GKM2012{[} MN30	&	\ldots			&	J15484207$-$5507422		&	16.52		&	7.93		&	6.73		&	5.98		&	\citet{2010AJ....139.2330W}\\
IRAS 16254-4739		&	\ldots			&	J16290377$-$4746264		&	\ldots		&	8.67		&	7.14		&	6.19		&	\citet{2010AJ....139.2330W}		\\
IRAS 16278-4808		&	{[}GKF2010{]} MN42 & J16313781$-$4814553	&	\ldots		&	\ldots		&	12.60		&	10.28		&	\citet{2012IAUS..282..267S} \\	
{\bf{EM* VRMF 55}}	&	MN44			&	J16323995$-$4942137		&	15.00		&	8.41		&	7.46		&	6.81		&	\citet{2015MNRAS.454.3710G} \\
HD 148937 			&	WRAY 19-46		&	J16335238$-$4806404		&	6.71		&	5.87		&	5.74		&	5.64		&	vG01		\\
{[}GKF2010{[} MN45 	& \ldots			&	J16364278$-$4656207		&	\ldots		&	11.71		&	7.89		&	5.82		&	\citet{2010AJ....139.2330W}			\\
{{\bf Cl* Westerlund 1 W 243}} & \ldots	&	J16470749$-$4552290		&	15.73		&	6.41		&	5.27		&	4.63		&	\citet{2009AaA...507.1597R}			\\
{\bf{{[}GKF2010{]} MN48}} &	 \ldots		&	J16493770$-$4535592		&	15.83		&	7.24		&	6.09		&	5.42		&	\citet{2016MNRAS.459.3068K} \\
$\zeta^1$ Sco 		& HD 152236 		&	J16535972$-$4221433		&	4.79		&	3.50		&	3.27		&	3.13		&	vG01 		\\
HD 326823 			&	V1104 Sco		&	J17065390$-$4236397		&	9.05		&	6.71		&	6.10		&	5.45		&	vG01 		\\
\ldots 				&	\ldots			&	J17082913$-$3925076		&	\ldots		&	10.49		&	8.22		&	7.11		&	\citet{2010AJ....139.2330W}			\\
{[}GKF2010{]} MN55	&	\ldots			&	J17110094$-$3945174		&	\ldots		&	9.73		&	8.04		&	6.96		&	\citet{2010AJ....139.2330W}			\\
{[}B61{]} 2 		&	WRAY 17-96		&	J17413543$-$3006389		&	15.00		&	6.71		&	5.52		&	4.80		&	C05			\\
{[}GKF2010{]} WS2	&	Hen 3-1383 		&	J17203175$-$3309490		&	14.57		&	6.62		&	5.41		&	4.70		& \citet{2012MNRAS.421.3325G}	\\
{\bf HD 160529}		&	 V905 Sco		&	J17415902$-$3330136		&	6.66		&	3.69		&	3.23		&	2.95		&	HD94		\\
{[}GKF2010{]} MN61	&	\ldots			&	J17421401$-$2955360		&	\ldots		&	13.38		&	10.54		&	8.99		&	\cite{2012ASPC..465..514S}			\\
{\bf{{[}GKF2010{]} MN58}}& \ldots		&	{1737476$-$313732}		&	\ldots		&	11.77		&	9.25		&	7.79		&	\cite{2012ASPC..465..514S} \\	
{\bf GCIRS 34W}		&	WR 101db		&	\ \ 17453973$-$2900266	&	\ldots		&	\ldots		&	\ldots		&	11.60		&	C05		\\
GCIRS 33SE			&	\ldots			& 	\ \ 17454002$-$2900310	&	\ldots		&	\ldots		&	\ldots		&	\ldots		&	C05			\\
GCIRS 16NW 			&	WR 101j			&	 17454005$-$2900269		&	\ldots		&	15.04		&	12.04		&	10.03		&	C05 		\\ 
GCIRS 16SW			&	\ldots			&	\ \ 17454012$-$2900291	&	\ldots		&	14.75		&	11.60		&	10.06		&	C05		\\ 
GCIRS 16C			&	WR 101l			&	\ \ 17454012$-$2900276	&	\ldots		&	15.23		&	11.99		&	9.83		&	C05		\\ 
GCIRS 16NE			&	\ldots			&	\ \ 17454026$-$2900272	&	\ldots		&	14.01		&	10.94		&	9.00		&	C05		\\
{\bf LBV G0.120-0.048}	&	{V4998 Sgr}	&	J17460562$-$2851319		&	\ldots		&	12.53		&	9.24		&	7.46		&	\citet{2010ApJ...713L..33M} 	\\
Pistol Star			&	V4647 Sgr		&	J17461524$-$2850035		&	\ldots		&	11.83		&	8.92		&	7.30		&	vG01		\\
{\bf FMM 362}		&	V4650 Sgr		&	J17461798$-$2849034		&	\ldots		&	12.31		&	8.97		&	7.09		&	C05			\\
WR 102ka			&	Peony Star		&	J17461811$-$2901366		&	\ldots		&	12.98		&	10.27		&	8.84		&	C05			\\
HD 316285 			&	V4375 Sgr		&	J17481403$-$2800531		&	9.60		&	4.82		&	4.23		&	3.58		&	vG01 		\\
{[}GKF2010{]} MN76	&	\ldots 			&	J18070516$-$2015163		&	\ldots		&	14.87		&	12.67		&	11.16		&	\citet{2012IAUS..282..267S} \\
{[}KMN95{]} Star A 	&	LBV 1806-20		&	J18084031$-$2024411		&	\ldots		&	13.66		&	10.48		&	8.74		&	C05			\\
{\bf HD 168607} 	&	V4029 Sgr 		&	J18211489$-$1622318		&	8.28		&	4.51		&	3.88		&	3.45		&	vG01		\\ 
HD 168625			&	V4030 Sgr		&	J18211955$-$1622260		&	8.37		&	5.07		&	4.54		&	4.18		&	vG01 		\\
{\bf MWC 930}		&	V446 Sct		&	J18262523$-$0713177		&	11.51		&	6.65		&	5.81		&	5.26		&	\citet{2014AdAst2014E...7M}		\\
{[}GKF2010{]} MN 80	&	\ldots			&	J18333954$-$0807084		&	\ldots		&	13.00		&	9.85		&	8.15		&	\citet{2012ASPC..465..514S}		\\
{\bf V481 Sct}		&	G24.73+0.69		&	J18335528$-$0658386		&	\ldots		&	8.36		&	6.84		&	5.92		&	C05		\\
{[}WBH2005{]} G025.520$+$0.216 & IRAS18344$-$0632 &	J18370520$-$0629379	&	\ldots	&	15.80		&	13.64		&	11.62		&	HD94	\\
BD-13 5061 			&	AS 314			&	J18392610$-$1350470		&	9.89		&	7.86		&	7.63		&	7.36		&	vG01	\\
{[}GKF2010{]} MN83	&	\ldots			&	J18392300$-$0553199		&	\ldots		&	13.19		&	10.09		&	8.39		& 	\citet{2012ASPC..465..514S}	\\
MSX6C G026.4700+00.0207	&	\ldots		&	J18393224$-$0544204		&	\ldots		&	8.00		&	6.53		&	5.61		&	C05		\\
{[}GKF2010{]} MN84 	&	\ldots			&	J18415965$-$0515409		&	\ldots 		&	7.96		&	6.53		&	5.68		& 	\citet{2010AJ....139.2330W}			\\
MGE G032.0204-00.3538 &	{[}GKF2010{]} MN96 & \ \ 1851021$-$005821	&	\ldots		&	\ldots		&	\ldots		&	\ldots		&	\citet{2012IAUS..282..267S} \\
WRAY 16-232 		&	 MN46			&	J16431636$-$4600424		&	12.5		&	6.26		&	5.08		&	4.21		&	\citet{2010MNRAS.405.1047G}		\\
{\bf IRAS 18576+0341} &	AFGL 2298		&	J19001089$+$0345471		&	\ldots		&	12.16		&	8.92		&	6.91		&	C05		\\
IRAS 19040+0817		&	 MN101			&	J19062457$+$0822015		&	\ldots		&	9.66		&	7.93		&	6.85		&	\citet{2012ASPC..465..514S}	\\
BD$+$14 3887 		&	MWC 314			&	J19213397$+$1452570		&	9.89		&	6.09		&	5.54		&	5.02		&	vG01 \\
{[}OMN2000{]} LS1	&	\ldots			&	J19234764$+$1436391		&	15.10		&	8.10		&	7.02		&	6.23		&	C05		\\
{[D75b] Em* 19-008} &	MN 112			&	J19443759$+$2419058		&	14.64		&	8.86		&	8.02		&	7.42		&	\citet{2010MNRAS.405..520G}		\\
{\bf P Cyg}			&	HD 193237		&	J20174719$+$3801585		&	4.82		&	3.79		&	3.51		&	3.27		&	HD94	\\
{MWC 1015} 			&	V439 Cyg		&	J20213356$+$3724515		&	11.84		&	8.53		&	8.01		&	7.59		&	\citet{1998AaA...339...75P}	\\
G 79.29+0.46		&	\ldots			&	\ \ 2031421$+$402201	&	\ldots		&	6.91		&	5.29		&	4.33		&	HD94\\
Cyg OB2 12			&	AS 424			&	J20324096$+$4114291		&	12.54		&	4.67		&	3.51		&	2.70		&	vG01	\\
MWC 349A			&	V1478 Cyg		&	J20324553$+$4039366		&	13.15		&	6.23		&	4.75		&	3.15		&	\citet{2012AaA...541A...7G}	\\
\hline \multicolumn{8}{c}{\bf Small Magellanic Cloud LBVs}\\ \hline
{LHA 115-S 6} 		& 	{R 4}			&	J00465501$-$7308342		&	12.97		&	12.17		&	11.73		&	11.03		&	vG01	 \\
LHA 115-S 18 		&	AzV 154			&	J00540955$-$7241431		&	13.82		&	12.35		&	11.93		&	11.11		&	\citet{2002AaA...386..926V} \\
{\bf HD 5980}		&	R 14			&	J00592656$-$7209540		&	11.31		&	11.11		&	11.01		&	10.77		&	\citet{2010AJ....139.2600K}	 \\
{\bf HD 6884} 		&	R 40			&	J01071821$-$7228035		&	10.20		&	9.71		&	9.53		&	9.47		&	vG01\\
\hline \multicolumn{8}{c}{\bf Large Magellanic Cloud LBVs}\\ \hline
HD 268835 			& 	{R 66} 			& 	J04564705$-$6950247 	& 	10.63 		& 	10.08 		& 	9.68 		& {8.86} 		& \citet{1983AaA...120..287S} \\
{\bf HD 269006}		&	R 71			&	J05020738$-$7120131		&	10.55 		&	10.67		&	10.64		&	10.55		&	HD94	\\
{HD 268939}			&	R 74			&	J05041492$-$6715052		&	{11.03}		&	10.60		&	10.51		&	10.36		&	vG01	\\
LHA 120-S 18 		& SK $-$68 42 		& 	J05055395$-$6810505 	& 	12.07 		& 	{11.66} 	& 	{11.57} 	& 	{11.39} 	&  	vG01 	\\
HD 269050			&	R 78			&	J05072041$-$6832085		&	11.54		&	11.74		&	11.75		&	11.75		&	vG01	\\
{HD 269128}			&	R 81			&	J05102280$-$6846238		&	{10.52}		&	10.24		&	10.14		&	10.09		&	vG01	\\
{\bf HD 269216} 	& SK $-$69 75 		&	J05133077$-$6932236		&	11.12		&	10.45		&	10.38		&	10.26		&	vG01	\\
{HD 34664} 			&	MWC 105			&	J05135297$-$6726548		&	{11.59}		&	10.37		&	9.75		&	8.46		&	vG01	\\
{HD 269227} 		&	R 84			&	J05135429$-$6931465		&	{12.71}		&	9.41		&	8.54		&	8.12		&	HD94	\\
{HD 269321}			&	R 85			&	J05175607$-$6916037		&	{10.84}		&	10.10		&	9.98		&	9.82		&	vG01	\\
{\bf S Dor}			&	HD 35343		&	J05181435$-$6915010		&	{10.25} 	&	8.68		&	8.51		&	8.34		&	HD94	\\
{HD 269445} 		&	R 99			&	J05225978$-$6801466		&	{11.51}		&	10.54		&	10.32		&	10.01		&	vG01	\\
{\bf HD 269582}		&	{SK $-$69 142a}	&	J05275266$-$6859084		&	{10.73}		&	{12.04}		&	{11.93}		&	{11.64}		&	vG01 	\\
{HD 269604} 		& SK $-$68 93 		& 	J05283137$-$6853557 	& 	{10.74} 	& 	{10.39} 	&	 {10.37} 	& 	{10.28} 	& 	vG01  \\
{\bf HD 269662}		&	 R 110			&	J05305147$-$6902587		&	{10.28}		&	9.78		&	9.67		&	9.50		&	HD94	\\
{HD 269687}			&	LHA 120-S 119	&	J05312554$-$6905384		&	{11.87}		&	11.82		&	11.85		&	11.72		&	vG01	\\
{\bf HD 269700} 	&	R 116			&	J05315227$-$6832388		&	10.54		&	10.46		&	10.37		&	10.32		&	vG01	\\
{HD 37836} 			&	R 123			&	J05351663$-$6940384		&	{10.69}		&	9.89		&	9.72		&	9.38		&	vG01	\\
{HD 37974}			&	{R 126}			&	{J05362586$-$6922558}	&	10.99		&	10.13		&	9.76		&	8.76		&	\citet{2002AaA...386..926V} \\
{\bf HD 269858}		&	R 127			&	J05364370$-$6929474		&	{10.15}		&	9.63		&	9.51		&	9.23		&	HD94	\\
{HD 269859}			&	{R 128}			&	J05364719$-$6929522		&	{10.73} 	& 	{10.84} 	& {10.89} 		& 	{10.84} 	& 	vG01	\\
{\bf {R143}}		&	CPD-69 463		& 	J05385162$-$6908071		&	{12.01} 	&	10.50		&	10.25		&	10.05		&	HD94	\\
{R 149} 			&	SK $-$69 257	&	J05395875$-$6944040		&	12.52		&	12.65		&	12.66		&	12.68		&	vG01	\\
{HD 38489} 			&	MWC 126			&	J05401333$-$6922464		&	{11.93}		&	10.46		&	9.85		&	8.60		&	vG01	\\
{CPD$-$69 500}		&	{SK $-$69 271}	&	{J05412014$-$6936229}	&	11.79		&	11.98		&	11.97		&	11.92		&	\citet{1997AaA...325.1157W} \\
SK $-$69 279		&	\ldots			&	J05414467$-$6935150		&	12.84		&	12.68		&	12.58		&	12.53		&	vG01	\\
LHA 120-S 61		&	SK $-$67 266 	& 	J05455192$-$6714259		&	{11.95}		&	12.03		&	12.00		&	11.90		&	HD94	\\
\hline \multicolumn{8}{c}{\bf M31 LBVs}\\ \hline
\ldots      		&	\ldots			&	003910.85$+$403622.4 	&	18.18		&	\ldots		&	\ldots		&	\ldots 		&	M16, H17	\\
{\bf LGGS J004051.59+403303.0}	& \ldots &	004051.59$+$403303.0	&	16.99		&	16.38		&	15.59		&	15.75		&	M07, \citet{2015MNRAS.447.2459S}	\\
{M31 V0571}        	&	\ldots			&	{004109.48$+$404852.0}	&	{20.50}		&	16.41		&	15.96		&	15.09		&	M07, H17	\\
{\bf AE And} 		&	HV 4476			&	J00430251$+$4149121		&	17.43		&	16.39		&	17.03		&	16.06		&	HD94	\\
\ldots        		&	\ldots			&	004322.47660$+$413940.8832	&	20.35	&	\ldots		&	\ldots		&	\ldots		&	M07		\\
{\bf AF And}		&	HV 4013			& 	J00433308$+$4112103		&	17.33		&	15.82		&	15.35		&	14.84		&	HD94	\\   
\ldots        		&	\ldots			&	J00434183$+$4111120		&	17.55		&	16.36		&	16.18		&	15.56		&	M07		\\
\ldots 				& \ldots			&	J00435048$+$4146110 	&	17.74		&	16.25		&	15.85		&	15.95		&	M07, \citet{2015MNRAS.447.2459S}		\\
{[}MLV92{]} 339869  &	{[}WB92a{]} 315A &	J00441132$+$4132568		&	18.07		&	16.60		&	16.40		&	15.49		&	M07, H17		\\
{\bf Var 15}		&	{[}WB92a{]} 370	&	J00441944$+$4122464		&	18.45		&	16.0  		& 	15.6 		& 	15.3		&	HD94		\\
{[}WB92a{]} 411 	& \ldots 			&	J00442520$+$4134519		&	17.48		&	16.74		&	16.15		&	16.90		&	HD94, H17		\\
\ldots 				&	\ldots 			&	004443.99$+$415151.7	&	19.03		&	\ldots 		&	\ldots 		&	\ldots 		&	M16, \citet{2015MNRAS.447.2459S}		\\
{\bf Var A-1}		&	\ldots			&	J00445054$+$4130372		&	17.14		&	15.75		&	15.54		&	15.46		&	HD94		\\
{MAC 2-123}       	&	\ldots			&	004507.66$+$413740.5	&	16.15		&	15.40		&	15.30		&	15.00		&	M07		\\
\ldots        		&	\ldots			&	J00452257$+$4150346		&	18.50		&	17.00		&	16.29		&	15.42		&	M07		\\
{\bf UCAC4 660-003111} &	\ldots		& 	J00452658$+$4150057		&	{16.39}		&	15.93		&	15.76		&	15.10		&	M07, \citet{2015MNRAS.447.2459S}		\\
\hline \multicolumn{8}{c}{\bf M33 LBVs}\\ \hline
\ldots				&	\ldots			&	013228.99$+$302819.3 	&	19.00		&	\ldots		&	\ldots		&	\ldots		&	M16		\\
\ldots				&	\ldots			&	013235.21$+$303017.4	&	18.01		&	\ldots		&	\ldots		&	\ldots		&	M07		\\
{[}HS80{]} B7		&	{UIT 008	}	&	013245.37$+$303858.2	&	17.61		& 	\ldots		&	\ldots		&	\ldots		&	H17\\
\ldots				&	\ldots			&	013248.26$+$303950.4 	&	17.25		&	\ldots		&	\ldots		&	\ldots		&	M07		\\
{[}HS80{]} B43		&	\ldots			&	013300.02$+$303332.4	&	18.32		&	\ldots		&	\ldots		&	\ldots		&	M07		\\
{[}HS80{]} B48 		&	\ldots			& 	{J01330304$+$3031014}	&	17.00		&	\ldots		&	\ldots		&	\ldots		&	M16		\\
FSZ 83        		&	\ldots			&	013317.01$+$305329.87	&	18.68		&	\ldots		&	\ldots		&	\ldots		&	M16		\\
\ldots				&	\ldots			&	013317.22$+$303201.6	&	\ldots		&	\ldots		&	\ldots		&	\ldots		&	H17\\
{[DMS93] NGC 595 6} &	\ldots			&	013332.60$+$304127.1 	&	18.99		&	\ldots		&	\ldots		&	\ldots		&	M07		\\
FSZ 126				&	\ldots			&	013334.11$+$304744.6	&	17.52		&	\ldots		&	\ldots		&	\ldots		&	H17 \\
{\bf Var C}			&	\ldots			&	013335.10$+$303600.3	&	16.43		&	16.60		&	16.25		&	15.70		&	HD94	\\
\ldots				&	\ldots			&	013337.31336$+$303328.8	&	\ldots		&	\ldots		&	\ldots		&	\ldots		&	H17 \\
{FSZ 163}			&	\ldots			&	J01333949+3045405		&	17.50		&	{16.78}		&	{17.15}		&	{15.30}		&	M07		\\
{[HS80] 110A} 		&	\ldots			& 	J01334122$+$3022369		&	16.29		&	{16.27}		&	15.70		&	{15.40}		&	M07	\\
{[MAP95] M 33 OB66 28} & \ldots			& 	013344.79$+$304432.4	&	18.15		&	\ldots		&	\ldots		&	\ldots		&	M16		\\
{\bf Var B} 		&	\ldots			&	J01334919$+$3038091		&	16.21		&	14.25		&	14.07		&	13.96		&	HD94	\\
{IFM-B 1079}     	&	\ldots			&	013351.46$+$304057.0	&	17.73		&	\ldots		&	\ldots		&	\ldots		&	M07	\\
{[MJ98] WR 98} 		&	\ldots 			&	{J01335238+3039095}		&	20.72		&	15.75		&	15.32		&	14.69		&	\citet{2009MNRAS.396L..21V} \\
{[MJ98] WR 106}		&	\ldots			&	 013354.85$+$303222.8	&	18.34		&	\ldots		&	\ldots		&	\ldots		&	H17 \\
{[HS80] B324}		&	\ldots			& 	{J01335593$+$3045304}	&	14.86		&	13.71		&	13.43		&	13.28		&	M07	\\
\ldots		        &	\ldots			&	013357.73$+$301714.2	&	17.39		&	\ldots		&	\ldots		&	\ldots		&	M07	\\
{[}HS80{]} B416     &	\ldots			&	013406.60$+$304147.6	&	16.08		&	{15.40}		&	{15.40}		&	{15.00}		&	M07	\\
{[}HS80{]} B393   	&	{[MJ98] WR 123}	&	013406.79$+$304727.0	&	17.20		&	\ldots		&	\ldots		&	\ldots		&	H17	\\
{\bf Var 83} 		&	\ldots			&	J01341090$+$3034373 	&	15.40		&	15.56		&	15.42		&	15.26		&	M07	\\
{[HS80] B517}  		&	\ldots			&	013416.07$+$303642.1	&	17.95		&	\ldots		&	\ldots		&	\ldots		&	M07	\\
{[HS80] B526} 		&	\ldots			&	013416.10$+$303344.9	&	17.12		&	\ldots		&	\ldots		&	\ldots		&	M07	\\
\ldots 				&	\ldots			&	013416.42$+$303120.7	&	17.14		&	\ldots		&	\ldots		&	\ldots		&	M07	\\
{\bf Var 2} 		&	Y Tri			&	013418.35$+$303836.8	&	18.22		&	\ldots		&	\ldots		&	\ldots		&	HD94		\\
{FSZ 458}        	&	\ldots			&	013422.88$+$304411.1	&	17.22		&	\ldots     	&	\ldots		&	\ldots		&	M07	\\
{Pul -3 120290} 	&	\ldots			&	JJ01342475$+$3033061	&	16.60		&	{16.34}		&	{16.80}		&	{16.30}		&	M07	\\
FSZ 465				&	\ldots			&	J01342718$+$3045599		&	18.50		&	16.38		&	15.72		&	15.46		&	H17 \\
\ldots				&	\ldots			&	J013429.63$+$303732.1	&	17.11		&	{16.80}		&	{16.05}		&	{15.80}		&	M07	\\
\ldots				&	\ldots			&	013432.76$+$304717.2	&	19.09		&	\ldots		&	\ldots		&	\ldots		&	H17 \\
{[}MAP95{]} M 33 OB88 7 &	\ldots		&	013459.36$+$304201.0	&	18.25		&	\ldots		&	\ldots		&	\ldots		&	H17 \\
\bf{Romano's Star}	&	M33 V0532		&	J01350971$+$3041565		&	18.04		&	16.83		&	16.70		&	16.80		& H17 \\
\hline \multicolumn{8}{c}{\bf IC10 LBVs}\\ \hline
\ldots				&		\ldots		&	002012.13$+$591848.0	&	19.37		&	\ldots		&	\ldots		&	\ldots		&	M07		\\
\ldots				&		\ldots		&	002016.48$+$591906.9	&	19.19		&	\ldots		&	\ldots		&	\ldots		&	M07			\\
\ldots				&		\ldots		&	002020.35$+$591837.6	&	19.11		&	\ldots		&	\ldots		&	\ldots		&	M07			\\
\hline \multicolumn{8}{c}{\bf {IC 1613 LBV}}\\ \hline
{[S71b] V39}		&	{\ldots}		&	J01050208$+$0210246		&	18.85		&	16.71		&	15.95		&	15.71		&	\citet{2010AaA...513A..70H}	\\	
\hline \multicolumn{8}{c}{\bf NGC 2366 LBV}\\ \hline
{\bf NGC 2363 V1}	&	{\ldots}		&	{072843.37$+$691123.9}	&	{17.88}		&	{15.88}		&	{14.88}		&	{14.12}		&	\citet{2006AJ....132.1756P}	\\	
\hline \multicolumn{8}{c}{\bf NGC 6822 LBV}\\ \hline
\ldots				&		\ldots		&	194503.77$-$145619.1	&	18.24		&	\ldots		&	\ldots		&	\ldots		&	M07		\\
\hline
\enddata
\tablecomments{The abbreviated references in the table are as follows: HD94 = \citet{HD94}; vG01 = \citet{vG01}; C05 = \citet{C05}; M07 = \citet{2007AJ....134.2474M}; M16 = \citet{2016AJ....152...62M}; H17 = \citet{2017ApJ...836...64H}. }
\end{deluxetable}  
\end{longrotatetable}

\acknowledgments

\bibliographystyle{aasjournal}
\bibliography{LBVs}

\begin{thebibliography}{}
\expandafter\ifx\csname natexlab\endcsname\relax\def\natexlab#1{#1}\fi
\providecommand{\url}[1]{\href{#1}{#1}}

\bibitem[{{Clark} {et~al.}(2005){Clark}, {Larionov}, \& {Arkharov}}]{C05}
{Clark}, J.~S., {Larionov}, V.~M., \& {Arkharov}, A. 2005, \aap, 435, 239

\bibitem[{{Davidson} \& {Humphreys}(2012)}]{2012ASSL..384.....D}
{Davidson}, K., \& {Humphreys}, R.~M., eds. 2012, Astrophysics and Space
  Science Library, Vol. 384, {Eta Carinae and the Supernova Impostors}

\bibitem[{{Gvaramadze} {et~al.}(2015){Gvaramadze}, {Kniazev}, \&
  {Berdnikov}}]{2015MNRAS.454.3710G}
{Gvaramadze}, V.~V., {Kniazev}, A.~Y., \& {Berdnikov}, L.~N. 2015, \mnras, 454,
  3710

\bibitem[{{Gvaramadze} {et~al.}(2014){Gvaramadze}, {Kniazev}, {Berdnikov},
  {Langer}, {Grebel}, \& {Bestenlehner}}]{2014MNRAS.445L..84G}
{Gvaramadze}, V.~V., {Kniazev}, A.~Y., {Berdnikov}, L.~N., {et~al.} 2014,
  \mnras, 445, L84

\bibitem[{{Gvaramadze} {et~al.}(2010{\natexlab{a}}){Gvaramadze}, {Kniazev}, \&
  {Fabrika}}]{2010MNRAS.405.1047G}
{Gvaramadze}, V.~V., {Kniazev}, A.~Y., \& {Fabrika}, S. 2010{\natexlab{a}},
  \mnras, 405, 1047

\bibitem[{{Gvaramadze} {et~al.}(2010{\natexlab{b}}){Gvaramadze}, {Kniazev},
  {Fabrika}, {Sholukhova}, {Berdnikov}, {Cherepashchuk}, \&
  {Zharova}}]{2010MNRAS.405..520G}
{Gvaramadze}, V.~V., {Kniazev}, A.~Y., {Fabrika}, S., {et~al.}
  2010{\natexlab{b}}, \mnras, 405, 520

\bibitem[{{Gvaramadze} \& {Menten}(2012)}]{2012AaA...541A...7G}
{Gvaramadze}, V.~V., \& {Menten}, K.~M. 2012, \aap, 541, A7

\bibitem[{{Gvaramadze} {et~al.}(2012){Gvaramadze}, {Kniazev}, {Miroshnichenko},
  {Berdnikov}, {Langer}, {Stringfellow}, {Todt}, {Hamann}, {Grebel}, {Buckley},
  {Crause}, {Crawford}, {Gulbis}, {Hettlage}, {Hooper}, {Husser}, {Kotze},
  {Loaring}, {Nordsieck}, {O'Donoghue}, {Pickering}, {Potter}, {Romero
  Colmenero}, {Vaisanen}, {Williams}, {Wolf}, {Reichart}, {Ivarsen}, {Haislip},
  {Nysewander}, \& {LaCluyze}}]{2012MNRAS.421.3325G}
{Gvaramadze}, V.~V., {Kniazev}, A.~Y., {Miroshnichenko}, A.~S., {et~al.} 2012,
  \mnras, 421, 3325

\bibitem[{{Herrero} {et~al.}(2010){Herrero}, {Garcia}, {Uytterhoeven},
  {Najarro}, {Lennon}, {Vink}, \& {Castro}}]{2010AaA...513A..70H}
{Herrero}, A., {Garcia}, M., {Uytterhoeven}, K., {et~al.} 2010, \aap, 513, A70

\bibitem[{{Humphreys} \& {Davidson}(1994)}]{HD94}
{Humphreys}, R.~M., \& {Davidson}, K. 1994, \pasp, 106, 1025

\bibitem[{{Humphreys} {et~al.}(2017){Humphreys}, {Gordon}, {Martin}, {Weis}, \&
  {Hahn}}]{2017ApJ...836...64H}
{Humphreys}, R.~M., {Gordon}, M.~S., {Martin}, J.~C., {Weis}, K., \& {Hahn}, D.
  2017, \apj, 836, 64

\bibitem[{{Humphreys} {et~al.}(2014){Humphreys}, {Weis}, {Davidson}, {Bomans},
  \& {Burggraf}}]{2014ApJ...790...48H}
{Humphreys}, R.~M., {Weis}, K., {Davidson}, K., {Bomans}, D.~J., \& {Burggraf},
  B. 2014, \apj, 790, 48

\bibitem[{{Humphreys} {et~al.}(2016){Humphreys}, {Weis}, {Davidson}, \&
  {Gordon}}]{2016ApJ...825...64H}
{Humphreys}, R.~M., {Weis}, K., {Davidson}, K., \& {Gordon}, M.~S. 2016, \apj,
  825, 64

\bibitem[{{Kniazev} {et~al.}(2015){Kniazev}, {Gvaramadze}, \&
  {Berdnikov}}]{2015MNRAS.449L..60K}
{Kniazev}, A.~Y., {Gvaramadze}, V.~V., \& {Berdnikov}, L.~N. 2015, \mnras, 449,
  L60

\bibitem[{{Kniazev} {et~al.}(2016){Kniazev}, {Gvaramadze}, \&
  {Berdnikov}}]{2016MNRAS.459.3068K}
---. 2016, \mnras, 459, 3068

\bibitem[{{Koenigsberger} {et~al.}(2010){Koenigsberger}, {Georgiev}, {Hillier},
  {Morrell}, {Barb{\'a}}, \& {Gamen}}]{2010AJ....139.2600K}
{Koenigsberger}, G., {Georgiev}, L., {Hillier}, D.~J., {et~al.} 2010, \aj, 139,
  2600

\bibitem[{{Massey} {et~al.}(2007){Massey}, {McNeill}, {Olsen}, {Hodge},
  {Blaha}, {Jacoby}, {Smith}, \& {Strong}}]{2007AJ....134.2474M}
{Massey}, P., {McNeill}, R.~T., {Olsen}, K.~A.~G., {et~al.} 2007, \aj, 134,
  2474

\bibitem[{{Massey} {et~al.}(2016){Massey}, {Neugent}, \&
  {Smart}}]{2016AJ....152...62M}
{Massey}, P., {Neugent}, K.~F., \& {Smart}, B.~M. 2016, \aj, 152, 62

\bibitem[{{Mauerhan} {et~al.}(2010){Mauerhan}, {Morris}, {Cotera}, {Dong},
  {Wang}, {Stolovy}, {Lang}, \& {Glass}}]{2010ApJ...713L..33M}
{Mauerhan}, J.~C., {Morris}, M.~R., {Cotera}, A., {et~al.} 2010, \apjl, 713,
  L33

\bibitem[{{Miroshnichenko} {et~al.}(2014){Miroshnichenko}, {Manset},
  {Zharikov}, {Zsarg{\'o}}, {Ju{\'a}rez Jim{\'e}nez}, {Groh}, {Levato},
  {Grosso}, {Rudy}, {Laag}, {Crawford}, {Puetter}, {Reichart}, {Ivarsen},
  {Haislip}, {Nysewander}, \& {LaCluyze}}]{2014AdAst2014E...7M}
{Miroshnichenko}, A.~S., {Manset}, N., {Zharikov}, S.~V., {et~al.} 2014,
  Advances in Astronomy, 2014, 130378

\bibitem[{{Petit} {et~al.}(2006){Petit}, {Drissen}, \&
  {Crowther}}]{2006AJ....132.1756P}
{Petit}, V., {Drissen}, L., \& {Crowther}, P.~A. 2006, \aj, 132, 1756

\bibitem[{{Polcaro} \& {Norci}(1998)}]{1998AaA...339...75P}
{Polcaro}, V.~F., \& {Norci}, L. 1998, \aap, 339, 75

\bibitem[{{Polcaro} {et~al.}(2016){Polcaro}, {Maryeva}, {Nesci}, {Calabresi},
  {Chieffi}, {Galleti}, {Gualandi}, {Haver}, {Mills}, {Osborn}, {Pasquali},
  {Rossi}, {Vasilyeva}, \& {Viotti}}]{2016AJ....151..149P}
{Polcaro}, V.~F., {Maryeva}, O., {Nesci}, R., {et~al.} 2016, \aj, 151, 149

\bibitem[{{Rafelski} {et~al.}(2007){Rafelski}, {Ghez}, {Hornstein}, {Lu}, \&
  {Morris}}]{2007ApJ...659.1241R}
{Rafelski}, M., {Ghez}, A.~M., {Hornstein}, S.~D., {Lu}, J.~R., \& {Morris}, M.
  2007, \apj, 659, 1241

\bibitem[{{Ritchie} {et~al.}(2009){Ritchie}, {Clark}, {Negueruela}, \&
  {Najarro}}]{2009AaA...507.1597R}
{Ritchie}, B.~W., {Clark}, J.~S., {Negueruela}, I., \& {Najarro}, F. 2009,
  \aap, 507, 1597

\bibitem[{{Sholukhova} {et~al.}(2015){Sholukhova}, {Bizyaev}, {Fabrika},
  {Sarkisyan}, {Malanushenko}, \& {Valeev}}]{2015MNRAS.447.2459S}
{Sholukhova}, O., {Bizyaev}, D., {Fabrika}, S., {et~al.} 2015, \mnras, 447,
  2459

\bibitem[{{Skrutskie} {et~al.}(2006){Skrutskie}, {Cutri}, {Stiening},
  {Weinberg}, {Schneider}, {Carpenter}, {Beichman}, {Capps}, {Chester},
  {Elias}, {Huchra}, {Liebert}, {Lonsdale}, {Monet}, {Price}, {Seitzer},
  {Jarrett}, {Kirkpatrick}, {Gizis}, {Howard}, {Evans}, {Fowler}, {Fullmer},
  {Hurt}, {Light}, {Kopan}, {Marsh}, {McCallon}, {Tam}, {Van Dyk}, \&
  {Wheelock}}]{2006AJ....131.1163S}
{Skrutskie}, M.~F., {Cutri}, R.~M., {Stiening}, R., {et~al.} 2006, \aj, 131,
  1163

\bibitem[{{Smith}(2016)}]{2016MNRAS.461.3353S}
{Smith}, N. 2016, \mnras, 461, 3353

\bibitem[{{Smith} {et~al.}(2018){Smith}, {Aghakhanloo}, {Murphy}, {Stassun},
  {Drout}, \& {Groh}}]{2018arXiv180503298S}
{Smith}, N., {Aghakhanloo}, M., {Murphy}, J.~W., {et~al.} 2018, ArXiv e-prints,
  arXiv:1805.03298

\bibitem[{{Smith} \& {Tombleson}(2015)}]{2015MNRAS.447..598S}
{Smith}, N., \& {Tombleson}, R. 2015, \mnras, 447, 598

\bibitem[{{Stahl} {et~al.}(1983){Stahl}, {Wolf}, {Zickgraf}, {Bastian}, {de
  Groot}, \& {Leitherer}}]{1983AaA...120..287S}
{Stahl}, O., {Wolf}, B., {Zickgraf}, F.~J., {et~al.} 1983, \aap, 120, 287

\bibitem[{{Stringfellow} {et~al.}(2012{\natexlab{a}}){Stringfellow},
  {Gvaramadze}, {Beletsky}, \& {Kniazev}}]{2012IAUS..282..267S}
{Stringfellow}, G.~S., {Gvaramadze}, V.~V., {Beletsky}, Y., \& {Kniazev}, A.~Y.
  2012{\natexlab{a}}, in IAU Symposium, Vol. 282, From Interacting Binaries to
  Exoplanets: Essential Modeling Tools, ed. M.~T. {Richards} \& I.~{Hubeny},
  267--268

\bibitem[{{Stringfellow} {et~al.}(2012{\natexlab{b}}){Stringfellow},
  {Gvaramadze}, {Beletsky}, \& {Kniazev}}]{2012ASPC..465..514S}
{Stringfellow}, G.~S., {Gvaramadze}, V.~V., {Beletsky}, Y., \& {Kniazev}, A.~Y.
  2012{\natexlab{b}}, in Astronomical Society of the Pacific Conference Series,
  Vol. 465, Proceedings of a Scientific Meeting in Honor of Anthony F. J.
  Moffat, ed. L.~{Drissen}, C.~{Robert}, N.~{St-Louis}, \& A.~F.~J. {Moffat},
  514

\bibitem[{{Valeev} {et~al.}(2009){Valeev}, {Sholukhova}, \&
  {Fabrika}}]{2009MNRAS.396L..21V}
{Valeev}, A.~F., {Sholukhova}, O., \& {Fabrika}, S. 2009, \mnras, 396, L21

\bibitem[{{van Genderen}(2001)}]{vG01}
{van Genderen}, A.~M. 2001, \aap, 366, 508

\bibitem[{{van Genderen} \& {Sterken}(2002)}]{2002AaA...386..926V}
{van Genderen}, A.~M., \& {Sterken}, C. 2002, \aap, 386, 926

\bibitem[{{Wachter} {et~al.}(2010){Wachter}, {Mauerhan}, {Van Dyk}, {Hoard},
  {Kafka}, \& {Morris}}]{2010AJ....139.2330W}
{Wachter}, S., {Mauerhan}, J.~C., {Van Dyk}, S.~D., {et~al.} 2010, \aj, 139,
  2330

\bibitem[{{Weis} {et~al.}(1997){Weis}, {Chu}, {Duschl}, \&
  {Bomans}}]{1997AaA...325.1157W}
{Weis}, K., {Chu}, Y.-H., {Duschl}, W.~J., \& {Bomans}, D.~J. 1997, \aap, 325,
  1157

\bibitem[{{Wenger} {et~al.}(2000){Wenger}, {Ochsenbein}, {Egret}, {Dubois},
  {Bonnarel}, {Borde}, {Genova}, {Jasniewicz}, {Lalo{\"e}}, {Lesteven}, \&
  {Monier}}]{2000AaAS..143....9W}
{Wenger}, M., {Ochsenbein}, F., {Egret}, D., {et~al.} 2000, \aaps, 143, 9

\end{thebibliography}

\end{document}